\documentclass[preprint,superscriptaddress,amsmath,amssymb,aps,pre,floatfix]{revtex4-1}
\usepackage[body={17cm,25cm},top={1.5cm}]{geometry}
\usepackage[english]{babel}
\usepackage[utf8]{inputenc}
\usepackage{amsmath,amssymb,mathenv}
\usepackage{graphicx}
\usepackage [T1] {fontenc}
\usepackage{color}
\usepackage{bm}
\usepackage{dutchcal}

\newcommand{\Norm}{\mathcal{N}}
\newcommand{\br}{{\bf r}}
\newcommand{\bp}{{\bf p}}
\newcommand{\bx}{{\bf x}}
\newcommand{\bX}{{\bf X}}
\newcommand{\bP}{{\bf P}}
\newcommand{\smeq}{ \! = \!}
\newcommand{\xo}{\mathcal{x}}
\newcommand{\po}{\mathcal{p}}

\newcommand{\Eo}{\mathcal{E}}

\newcommand{\x}{x}

\usepackage{lmodern}
\usepackage{microtype}
\usepackage{hyperref}
\usepackage{amsfonts, vmargin}
\usepackage{tikz}
\usepackage{epic, eepic}
\usepackage[us]{datetime}

\usepackage[notref,notcite]{}
\definecolor{BLACK}{named}{black}
\definecolor{GREEN}{named}{green}

\newcommand{\chkout}[1]{\unskip}

\begin{document}
\title{Mean Field Games in the weak noise limit : A WKB approach to the Fokker-Planck equation}
\author{Thibault Bonnemain}
\affiliation{LPTMS, CNRS, Univ. Paris-Sud, Universit\'e Paris-Saclay,
  91405 Orsay, France}  
\affiliation{Laboratoire de Physique Théorique et Modélisation (CNRS UMR 8089), Universit\'e  de Cergy-Pontoise, F-95302
  Cergy-Pontoise, France}
\author{Denis Ullmo}
 \affiliation{LPTMS, CNRS, Univ. Paris-Sud, Universit\'e Paris-Saclay,
  91405 Orsay, France}  
\begin{abstract}
  { Motivated by the study of a Mean Field Game toy model called the
    ``seminar problem'', we consider the Fokker-Planck equation in the
    small noise regime for a specific drift field. This gives us the
    opportunity to discuss the application to diffusion problem of the
    WKB approach ``\`a la Maslov \cite{Maslov81}'', making it possible
    to solve directly the time dependant problem in an especially
    transparent way.}
\end{abstract}
\date{\today}
\maketitle

\section{Introduction}

Mean Field Games \cite{LasryLions2006-1,LasryLions2006-2,Huang2006}
are characterized by the coupling between a forward
diffusion process for a density $m(\bx,t)$ of agents with state variable $\bx \in
\mathbb{R}^n$ at time $t$,  and a backward optimisation process
characterized by a value function $u(\bx,t)$. 
In the simple case of quadratic mean field games
\cite{Ullmo-PhyRep-2018} this 
takes the form of a system of coupled (forward) Fokker-Planck  and (backward) Hamilton-Jacobi-Bellman equations
\begin{align}
    &\partial_t m(\bx,t)  + \pmb \nabla . (m(\bx,t) \pmb
    {\bf a}(\bx,t))  -
    \frac{\sigma^2}{2}\Delta\, m(\bx,t) =0 \quad \mbox{FP} \; , \label{eq:FP}
    \\
    & \partial_t u(\bx,t)  - \frac{1}{2\mu} \|\pmb \nabla u(\bx,t)\|^2  + 
    \frac{\sigma^2}{2} \Delta \, u(\bx,t) =\tilde V[m_t](\bx)
    \quad \mbox{HJB} \; , \label{eq:HJB}
\end{align}
with initial and final conditions $m(\cdot,t \!=\!0) = m_0(\cdot)$,
$u(\cdot,T)=c_T(\cdot)$.  The coupling between the two PDE's is
provided by the right hand side of Eq.~\eqref{eq:HJB} which involves
the functional of the density $m$ at time $t$, $V[m_t](\bx)$, (which may
also have an explicit dependence in $\bx$), and by the fact that the
drift velocity in Eq.~\eqref{eq:FP} is given in term of the gradiant
of the value function as ${\bf a}(\bx,t) = - \frac{1}{\mu} \nabla
u(\bx,t)$.

In the noiseless limit $\sigma = 0$,  this system of equations reduces to a
transport equation coupled to a Hamilton-Jacobi equation, both of which we
associate with the classical dynamics of point particles.  This limit
is therefore rather intuitive, and in some respects simpler to analyse
than the noisy regime.  It turns out however that in many
circumstances this limit is ill defined, which implies that it is
mandatory to include a small but non zero noise.  In that case, what
one needs to analyse is the small (but non-zero) $\sigma$ limit of the system
Eqs.~\eqref{eq:FP}-\eqref{eq:HJB}, which quite naturally one would wish to
study in terms of ``classical trajectories'' to make contact with the
intuitive description one has in mind for the $\sigma=0$ limit.

To avoid any misunderstanding, we stress right away that in this paper
we will provide only a very modest step toward the solution of this
general problem.  To start with we will limit ourselves to the analysis
of the particular case of a specific Mean Field Game toy model, the
``seminar problem'', introduced by Gu\'eant and co worker in
\cite{GueantLasryLions2010}, and analysed in some details in
\cite{Swiecicki-PhysicaA-2016}.  This Mean Field Game problem consists
in finding the effective starting time of a seminar, fixed by a quorum
condition, when all the participants try to optimise their behaviour
to avoid arriving too late or too early.  The ``state variable'' $x$
is therefore one dimensional, and correspond simply to the physical
space in which the motion of the agents takes place (the corridor
leading to the seminar room) modeled as the negative real line $x \in
]-\infty,0]$, and absorbing boundary conditions are assumed at $x=0$
(since no one is expected to exit the seminar room).  Furthermore, the
functional $V[m_t](x)$ is taken uniformly zero, and the coupling
between the HJB equation and the density of agents is just provided by
a quorum condition on the number of agents in the room at the
beginning of the seminar.

\begin{figure} [h!]
	\includegraphics[scale=0.9]{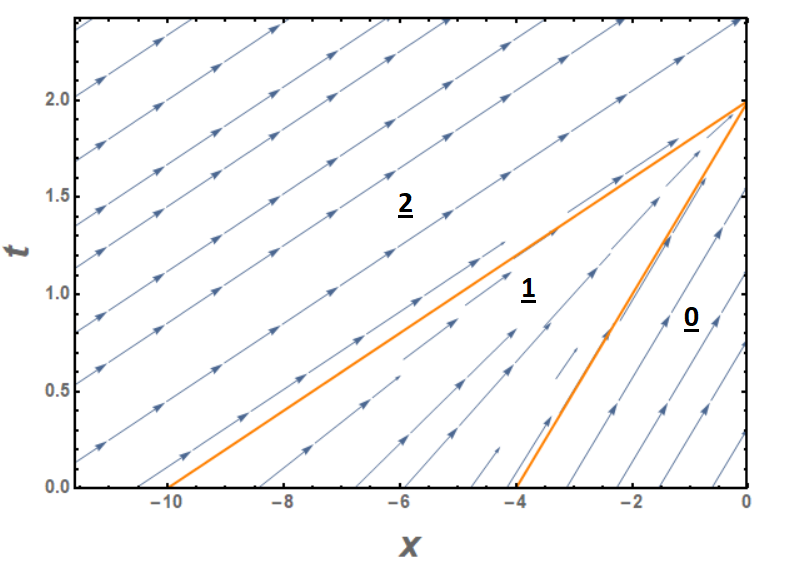}
	\caption{Regions of the $(t,{\bf x})$ space, where $T=2$ is the
          time when the seminar effectively begins, and their
          associated optimal drift $a(t,\bx)$. In regions (0) and (2)
          the drift stays constant and is denoted respectively $a^{(0)}$
          and $a^{(2)}$ (here 5 and 2). In region (1),the drift is linear in x.} 
	\label{fig:drift}
\end{figure}

In the weak noise regime, it is shown in
ref.~\cite{Swiecicki-PhysicaA-2016} that this problem is associated
with the drift field shown in Fig.~\ref{fig:drift}, and reading to
leading order 
\begin{equation} \label{eq:a_small_sigma}
a(t,x) = \left\{ 
\begin{aligned}
	  a^{(0)}  &  \quad \mbox{for } \quad & x \leqslant - a^{(0)} (T-t)  \\
	  \frac{-x}{(T-t)} & \quad \mbox{for } \quad & - a^{(0)} \leqslant x \leqslant
          - a^{(2)} (T-t) \\
	   a^{(2)} & \quad \mbox{for } \quad & -a^{(2)} (T-t)
         \leqslant x \leqslant  0
  \end{aligned} \right.    \; ,
\end{equation}
where $a^{(0)} > a^{(2)}$ are two constant drift velocities. 

The (admittedly limited) goal of this paper will therefore be to
analyse the Fokker-Planck equation for this velocity field in the
small $\sigma$ regime, and to show that we can provide a very precise
solution of this problem based just on the ``classical trajectories''
for a dynamics closely related to (but slightly different from) the
$\sigma=0$ limit of Eq.~\eqref{eq:FP}.  

The fact that this can be achieved for  the Fokker-Planck
equation can be seen readily  by multiplying Eq.~\eqref{eq:FP} by $\sigma^2$,
and noting that it then has the structure of what Maslov
\cite{Maslov81} has termed a ``$\lambda$-pseudo differential operator'',
in the sense that each partial derivative is associated with a factor
$\lambda^{-1} \equiv \sigma^2$.  This implies that a ``semiclassical
approximation'' scheme 
can be applied to this equation in small $\sigma^2$ limit.  This fact
has of course been recognized for many years, and led to some publications \cite{Risken,SHIZGAL90,Derevyanko:05}.  Most of them, however, use a
rather indirect approach, making use of transformation of
variable to a form more directly related to the Schr\"odinger equation
and through a normal mode decomposition (cf eg \cite{Caroli79} on
the example of a  diffusion in bistable potentials). We follow however
here the philosophy of the ray method introduced in \cite{CohenLewis67}.

Our goal will thus be to to show that a direct approach where the
time-dependent WKB approximation is applied directly on
Eq.~\eqref{eq:FP} can be used effectively to obtain a extremelly good
approximation for the solution of the Fokker-Planck equation
Eq.~\eqref{eq:FP} with the drift field \eqref{eq:a_small_sigma}.  We
address thus here only the first (and simplest) step of the analysis
of the coupled MFG equations system, and furthermore do this on a
specific illustrative case.  This gives us however the opportunity to
discuss the application of the WKB approach in the perspective
developed by Maslov \cite{Maslov81}, in a way which is maybe a bit
more transparent that what can be found in the literature
\cite{CohenLewis67}, and leads in our view to a rather
intuitive interpretation.

The paper will be organised as follows. In section~\ref{sec:recipe},
we will give without justification the recipe for the construction of
the WKB approximation.  For the sake of clarity this will be done for a
one dimensional problem, and we will assume that the initial density
$m_0(x)$ is a gaussian.  Section~\ref{sec:proof} will then provide a
derivation of these WKB expressions, together with a generalization to
higher dimensionality and to a larger class of initial densities.
Readers with little interest in these formal issues may skip that
section and go directly to section~\ref{sec:ill} where the WKB
approximation is applied to two simple examples where it turns out to
provide the exact solution, as well as to the case corresponding to
the drift field Eq.~\eqref{eq:a_small_sigma}.  Finally, we conclude in
section~\ref{sec:conclusion}, and, for self-containedness, briefly
sketch two rather standard derivations in appendices
\ref{app:characteristics} and \ref{app:liouville}.

\section{WKB approximation of a 1d Fokker-Planck equation}
\label{sec:recipe}

In this section, we provide, without any demonstration, the
prescription for the construction of  the WKB solution of the Fokker-Planck equation
Eq.~\eqref{eq:FP} in the small $\sigma$ regime. We limit
ourselves here to the one-dimensional case and to gaussian initial
densities
\begin{equation} \label{eq:IC}
	m_{0}(x)  = \Norm \exp\left[-\frac{\mu (x-\bar
            x_0)^2}{2\sigma^2}\right] \; ,
\end{equation}
where $\bar x_0$ is the center of the gaussian and $\Norm =
\sqrt{\frac{\mu}{2\pi\sigma^2}}$ is a normalisation factor.  More
general $m_0(x)$ could easily be considered (see
section~\ref{sec:proof}), but gaussians have an intrinsic interest, and,
in addition, this also allows us to get the Green's function of the
equation by reducing the width of the gaussian to zero.

The semiclassical scheme follows three steps. The first one consists in
constructing a Lagrangian symplectic manifold on which we can define
an action. The second step uses this input to
build the WKB approximation. Finally, we address how absorbing
boundary conditions can be implemented in the semiclassical scheme.

\subsection{Symplectic manifold and classical action}

The Fokker-Planck equation \eqref{eq:FP} can be written as $\hat L m
=0$ where we have introduced the $\lambda$-pseudo differential operator
$\hat L \equiv [\lambda^{-1} \partial_t \cdot +
\lambda^{-1} \partial_x(a \cdot) - \frac{1}{2}
(\lambda^{-1} \partial_x)^2 \cdot]$ (with again $\lambda \equiv
\sigma^{-2}$ assumed large).  Using the usual mapping
$\lambda^{-1} \partial_x \to p$, $\lambda^{-1} \partial_t \to E$,
$\hat L$ can be associated with the {\em classical symbol}
\begin{equation} \label{eq:L1}
	L(x,t;p,E) = E + pa(x,t) - p^2/2  \; ,
\end{equation}
which, if understood as a classical Hamiltonian 
leads to the  canonical equations 
\begin{equation} \label{eq:canon}
	\left\{
	\begin{aligned}
		&\dot t = \partial_E L = 1 \qquad & \dot E =
                - \partial_t L = -p\partial_t a\\ 
		&\dot x = \partial_p L = a(t,x) - p \qquad &\dot p =
                - \partial_x L = -p\partial_xa 
	\end{aligned}
	\right. \; .
\end{equation}

Now, consider the initial gaussian distribution Eq.~\eqref{eq:IC} for
$\bar x_0$ and $\mu$ given.  It can be written in the semiclassical form
$m_0(x_0) = \Norm \exp\left[ \lambda S_0(x_0)\right]$ with
\begin{equation} 
       S_0(x_0) \equiv - \mu \frac{ (x_0-\bar x_0)^2}{2} \; .
\end{equation}
At any
point of space $x_0$, one can therefore initiate a classical trajectory
at $t=t_0$ with an inital momentum 
\begin{equation} \label{eq:poxi}
  p_0(x_0) = \nabla S_0(x_0)= -\mu (x_0-\bar x_0) \; ,
\end{equation}
and fulfilling the ``compatibility condition''
\begin{equation} \label{eq:compatibility}
L(x,t;p,E)  \equiv 0 \; .
\end{equation}
The reunion of all these trajectories obtained from these intial
conditions and the canonical equations \eqref{eq:canon} form a
2-dimensional manifold  $\mathcal{M}= \{ (t,\xo(t,x_0),\Eo(t,x_0) \po(t,x_0) \}$ where $\po$, $\xo$ and $\Eo$ respectively represent the value taken by $p$, $x$ and $E$ after evolving on this manifold from $\br_0=(t_0, x_0;E_0(x_0),p_0(x_0))$ for a time $t-t_0$.

\begin{figure} [h!]
	\includegraphics[scale=0.5]{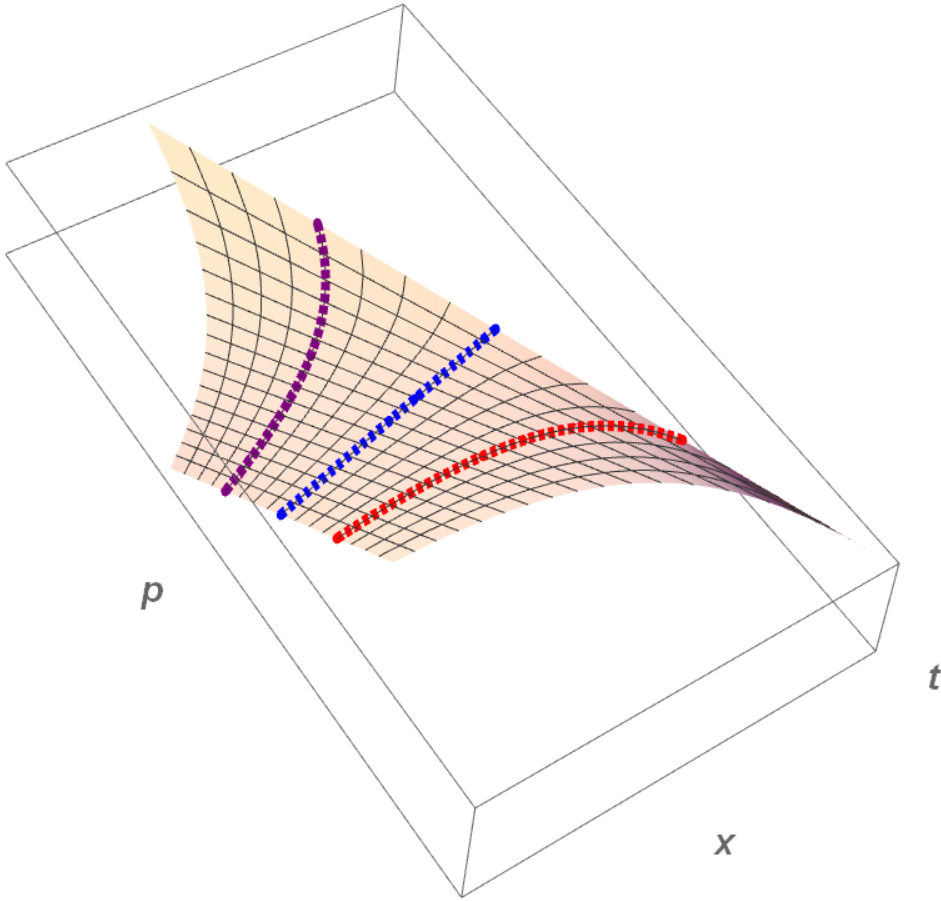}
	\caption{A typical manifold generated by the classical
          trajectories in region (1) of the drift field. In this case $a=\frac{x}{t-T}$, $\mu = 1.5$, $\sigma=0.4$, $T=2$ and $\bar x_0 =-5$. The dashed curves represent specific trajectories beginning at $x_0=-5.5$, $-5$ and $-4.5$ from left to right.} 
	\label{fig:manifold} 
\end{figure}

To the manifold  $\mathcal{M}$, we can now associate a classical
action 
\begin{equation} \label{eq:action}
       S(t,x) \equiv \int_{[\mathcal{L}: \bar \br_0 \to \br] \subset
         \mathcal{M}}  p dx + E dt \;  
\end{equation}
where $\bar \br_0 = (t_0,\bar x_0; E \smeq 0, p \smeq 0)$ is the point on
$\mathcal{M}$ above $\bar \bX_0 = (t_0,\bar x_0)$, and $\br \in \mathcal{M}$ is
the point above $\bX=(t,x)$.  

\begin{figure} [h!]
	\includegraphics[scale=0.58]{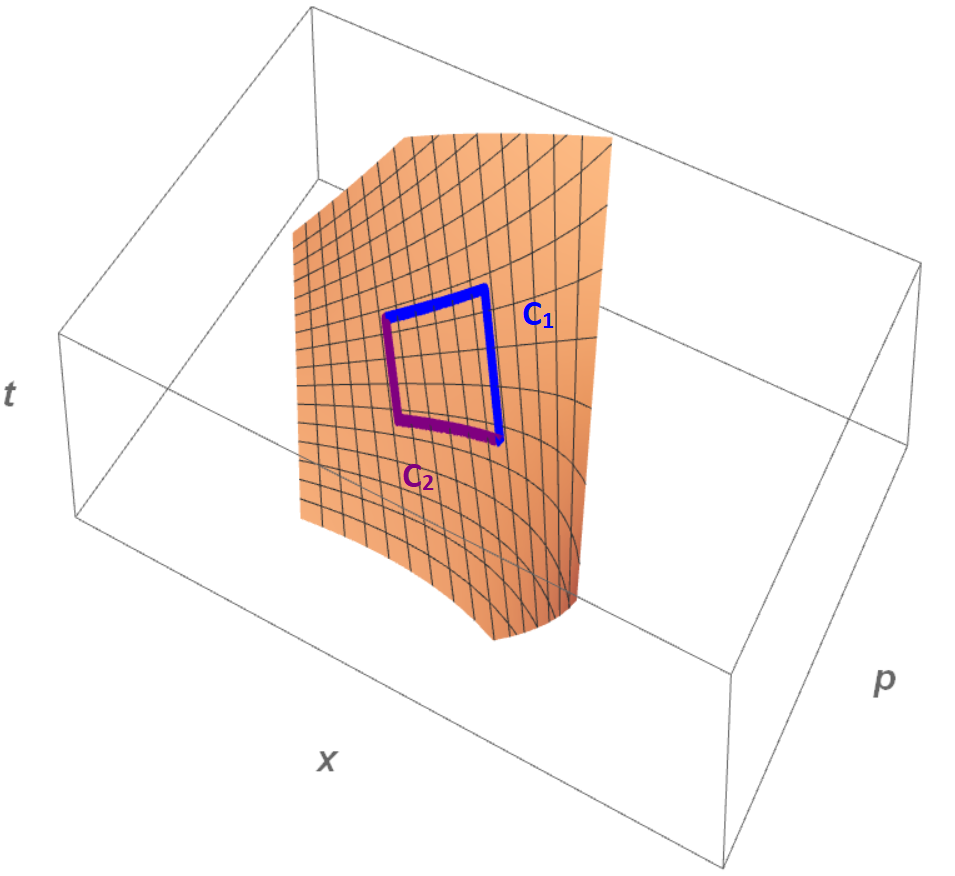}
	\caption{Same manifold as in Figure~\ref{fig:manifold} where are highlighted two paths with same beginning and end. Because of the Langrangian nature of the manifold we
          can write $\int_{C_1} (E\dot t + p\dot x) dt =\int_{C_2}
          (E\dot t + p\dot x) dt$.} 
	\label{fig:lag}
\end{figure}

We stress that, since
$\mathcal{M}$ is a Lagrangian manifold, the integral in
Eq.~\eqref{eq:action} can be taken on {\em any path} on $\mathcal{M}$
joining $\bar \br_0$ to $\br$. For instance,  the action $S(t,x)$ can be
computed either as 
\[ 
S_1(t,x) = \underbrace{\int_{\bar x_0}^{x_0(t,x)} p_0(x') dx'}_{S_0(x_0)} +
\int_{t_0}^t (\po(s,x_0) \dot \xo 
(s,x_0) + \Eo(s,x_0)) ds \; ,
\]
in which  $x_0(t,x)$ is the initial
position of the trajectory arriving at $x$ at time $t$, or as
\[ 
S_2(t,x) =  \int_{t_0}^t (\po(s,\bar x_0) \dot \xo
(s,\bar x_0) + \Eo(s,\bar x_0)) ds +
\int_{\xo(t,\bar x_0)}^{x} p(x',t) dx' \; ,
\]
with $p(x,t)$ the momentum coordinate of the point of
$\mathcal{M}$ above $(t,x)$.
Both expressions lead to the same result (i.e.\ $S_1(t,x) = S_2(t,x) =
S(t,x) $). This is illustrated on Fig.~\ref{fig:lag}.

For the gaussian initial density we consider, the definition of the initial momentum given by \eqref{eq:poxi} and the compatibility conditions  $L \equiv 0$ impose that $\po(t,\bar x_0) =0$ and $\Eo(t,\bar x_0) =0$  for all time, yielding
\begin{equation}
S(t,x) = S_2(t,x)= \int^{x}_{\xo(t,\bar x_0)} p(t,x')dx'\; , 
\end{equation}
where the path of integration on the manifold is taken at {\em
	constant time $t$} from the point above $\bar{x} \equiv
\xo(t,\bar x_0)$ (evolution of the center of the distribution
$\bar x_0$) to the point above $x$.

As a final comment, it is worth mentionning that, for more general
initial conditions, $S_0(\bar x_0)$ can be non-zero and should be
added to the right-hand side of \eqref{eq:action}.

\subsection{Semiclassical approximation for  \texorpdfstring{$m(t,x)$}{m(t,x)}}

With this definition of the action, the WKB approximation for the
density of probability is expressed as 
\begin{equation} \label{eq:mv1}
	m_{\rm s.c.}(t,x) = \frac{\Norm}{\sqrt{\partial_{x_0}\xo(t,x_0)}} 
 \exp \left[ \lambda S(t,x) -
   \frac{1}{2}\int^t_0(\partial_xa)d\tau \right] \; ,
\end{equation}
where in the prefactor, $\xo(t,x_0)$ is the position of a trajectory
started at $x_0$ at time $t \smeq t_0$ (with thus  a momentum $p_0(x_0)$
given by Eq.~\eqref{eq:poxi}), and the integral in the exponential is taken along
this trajectory.  Except for the fact that the exponent is real rather
than complex, the only difference with respect to the traditional WKB
expressions derived in optics or in the context of the Schr\"odinger
equation is the extra term $\frac{1}{2}\int^t_0(\partial_xa)d\tau$ in
the exponent, which can be tracked back to the non-symmetric
ordering of the operators $\hat p \equiv \lambda^{-1} \partial_x$ and
$\hat x \equiv \times x$ in the Fokker-Planck equation.

\subsection{Absorbing boundary conditions}
\label{sec:absorbing}

We shall illustrate below this WKB approach with the problem
corresponding to the drift velocity field
Eq.~\eqref{eq:a_small_sigma}, problem for which we assume an absorbing
boundary condition at $x=0$. As such absorbing boundary conditions are
rather common, we discuss now how to implement them in our
semiclassical scheme.

Let us consider the semiclassical solution of the free problem (ie
without the boundary condition)
\begin{equation}
          m_{\rm free}(t,x) =\frac{\Norm}{\sqrt{\partial_{x_0}\xo(t,x_0)}}
          \exp\left[\lambda S(t,x)  -
   \frac{1}{2}\int^t_0(\partial_xa)d\tau \right] \; .
\end{equation}
For sake of simplicity, we assume that (as will be the case in the
examples we are going to consider), the trajectories on which $S(t,x)$
is constructed are reaching $x=0$ with positive velocity.

Consider now the compatibility condition Eq.~\eqref{eq:compatibility}
at $x=0$,  for an arbitrary time $t$, and with the choice $E
= \partial_t S$
\[
L(0,t;p,\partial_t S) =0\; .
\]
  It admits two solutions
\begin{equation} 
\dot x =  a(x,t) - p = \pm \sqrt{a^2 + 2 \partial_t S} \; .
\end{equation}
The one corresponding to a positive velocity is just  $p_+(t)
= \partial_x S(t,x \smeq 0)$. We can however generate another set of
trajectories initiated at time $t$ at $x=0$ with momentum  $p_-(t) = a
+ \sqrt{a^2(t,0) + 2 \partial_t S(t,0)}$ and energy
$E(t) = \partial_t S(t,0)$.  These trajectories have negative
velocities and thus ``bounce'' off the boundary point $x=0$.

A ``reflected'' density 
\begin{equation}
          m_{\rm ref}(t,x) =\frac{\Norm}{\sqrt{\partial_{x_0} \tilde \xo(t,x_0)}}
          \exp\left[\lambda \tilde S(t,x) -
          \frac{1}{2}\int^t_0(\partial_xa)d\tau\right] \; ,
\end{equation}
can therefore be constructed in exactly the same way as before using
the reflected trajectories $\tilde\xo$ and reflected action $\tilde S$.  At $x=0$, $m_{\rm ref}(t,0) = m_{\rm
  free}(t,0)$ since $\partial_t S(t,0) = \partial_t \tilde S(t,0) =
E(t,0)$ is the same for both (and thus one should just impose $
S(t_0,0) = \tilde S(t_0,0)$ for an arbitrary time $t_0$), and
$\partial_{x_0}\tilde \xo(t,x_0) = \partial_{x_0} \xo(t,x_0)$ since at $x=0$
only the momentum has changed but not the position.  Therefore, the
total density
\begin{equation}
      m_{\rm tot}(t,x)    = m_{\rm free}(t,x) -  m_{\rm ref}(t,x) 
\end{equation}
is a semicalssical solution to the Fokker-Planck equation
\eqref{eq:FP} which fulfills the absorbing boundary condition $ m_{\rm
  tot}(t,0) = 0$.

\section{Derivation and generalisation}
\label{sec:proof}

We provide now a derivation (and some generalisation) of
Eq.~\eqref{eq:mv1}.  Our approach is very similar in spirit to the
``ray method'' developed by Cohen and Lewis \cite{CohenLewis67}, but
follow more closely the WKB formalism developped by Maslov
\cite{Maslov81}, that we feel might be easier to access for
physicists.  

We therefore want to describe the evolution of an initial density (at
$t=t_0$) which is in the ``semiclassical form''
\begin{equation} \label{eq:m0}
m_0(\bx) = \phi_0(\bx) \exp \left[\lambda S_0(\bx) \right] \; ,
\end{equation}
with $\bf x \in \mathbb{R}^d$.  Such form includes Gaussian densities
such as Eq.~\eqref{eq:IC}, but
are significantly more general.\\

By writing $(\sigma^{-2} \equiv \lambda)$,
the Fokker-Planck equation reads  in the more general case,
\begin{equation} \label{eq:FKSC}
	0=\lambda^{-1} \partial_t m + \lambda^{-1}\nabla ({\bf a}(t,\bx)m) -
        \frac{1}{2}\lambda^{-2}\Delta m= \hat L m \; ,
\end{equation}
which up to the $i$ factors, looks very much like a $\lambda$-pseudo
differential Maslov operator of symbol
\begin{equation} \label{eq:L-gen}
  L(\bx,t;\bp,E) = E +
  {\bf a}(\bx,t) \! \cdot \! \bp  - \bp^2/2 \; .
\end{equation}
Following Maslov's derivation \cite{Maslov81}, let us consider the ansatz
\begin{equation}
	m(t,\bx) =
        \phi(t,\bx)\exp \left[\lambda S(t,\bx) \right] \; ,
\end{equation}
with $\phi(t_0,\bx) = \phi_0(\bx)$ and $S(t_0,\bx) = S_0(\bx)$.\\

Writing $\bX \equiv (t,\bx)$, $\bP \equiv (E,\bp)$,
Eq.~\eqref{eq:FKSC} becomes
\begin{equation}
	\hat L \left[ \phi(\bX) e^{\lambda S(\bX)} \right]
        = 0 = e^{\lambda S(\bX)} \left[ R_0\phi(\bX) + \lambda^{-1} R_1\phi(\bX)
        + O(\lambda^{-2})\right] 
\end{equation}
with
\begin{equation}
	R_0 = L(\bX;\partial_{\bX} S) \; ,
\end{equation}
\begin{equation}
  R_1 = \langle \partial_{\bP}
  L(\bX;\partial_{\bX} S), \partial_{\bX}
  \phi \rangle +
  \frac{1}{2} \left\{
  {\rm Tr}\left[\partial^2_{\bP \bP} L(\bX; \partial_{\bX}S) 
  \partial^2_{\bX\bX}  S\right]     
  +
  {\rm Tr}\left[ \partial^2_{\bX \bP} L(\bX;\partial_{\bX} S)\right] \right\} \phi
\; .
\end{equation}
Neglecting terms of order $\lambda^{-2}$ and higher, solving
Eq.~\eqref{eq:FKSC} amounts to solving $R_0=0$ and $R_1=0$. 

\subsection{\texorpdfstring{$R_0=0$}{R0=0}, Hamilton-Jacobi equation} 

The equation $(R_0=0)$ can be rewritten as an Hamilton-Jacobi equation
on $S$ 
\begin{equation}
	L(\bX;\partial_{\bX} S) = \partial_tS + {\bf a}(t,\bx) \!
        \cdot \!  {\bf \nabla }
        S - \frac{1}{2}(\nabla S)^2 = 0  
	\label{eq:HJ} \; ,
\end{equation}
with an initial condition at $t=t_0$ 
\begin{equation}
 S(t_0,\bx_0) =  S_0(\bx_0) \; .
\end{equation}
Solution of this kind of equations is typically obtained through the
method of characteristics. Here this amounts to build a one paramater
 family of rays $(t, \bx; E,
\bp)_{\bx_0}(s)\equiv\br(s,\bx_0)$, indexed by $\bx_0$, which follow --
for a fictitious time $s$ -- the
Hamilton dynamics associated with $L$:
\begin{equation} \label{eq:Hdyn}
\left\{
  \begin{aligned}
    &\dot t = \partial_E L = 1 \qquad & \dot E = \partial_t L = -\bp \! \cdot \! \partial_t {\bf a} \  \  \ \\
    &\dot \bx = \partial_\bp L = {\bf a}(t,\bx) - \bp \qquad & \dot \bp = - \partial_\bx L = -\bp  \! \cdot \! \partial_\bx 
    {\bf a}\\    \end{aligned}
\right. \; ,
\end{equation}
with initial the conditions 
\begin{equation} \label{eq:rays}
  \left\{
    \begin{aligned}
      &\br(0,\bx_0) = (E_0, t_0,
      \bp_0(\bx_0), \bx_0) \\ 
      &L(\br(0,\bx_0))=0 \\
    \end{aligned}
  \right. \; 
\end{equation}
corresponding to 
\begin{equation}
	\bp_0(\bx_0) = \partial_{\bx_0} S_0(\bx_0) \; .
\end{equation}
Eq.~\eqref{eq:rays} fixes $E_0$ and it is clear from Eqs.~\eqref{eq:Hdyn} that we can take $s\equiv t-t_0$.

As stressed in the previous section, the family of rays defined by
Eqs.~\eqref{eq:Hdyn}-\eqref{eq:rays} form a Lagrangian manifold, thus,
according to the method of characteristics (cf
appendix~\ref{app:characteristics}), the solution of Eq.~\eqref{eq:HJ}
reads
\begin{equation} \label{eq:action-bis}
S(t,\bx) = \int_{\bar \bX_0}^{\bX} E dt + \bp \! \cdot \! d\bx \; ,
\end{equation}
where the integral is taken on any path on the
manifold starting above the point $ \bar \bX_0 = (t_0,\bar\bx_0)$ such that
$S_0(\bar \bx_0)=0$ and ending on the point above $\bX=(t,\bx)$.

\subsection{\texorpdfstring{$R_1=0$}{R1 = 0}, transport equation} 

We begin by focusing on the first term of $R_1$ that we rewrite more explicitly using the canonical Hamilton-Jacobi equations
\begin{equation}\langle \partial_{\bP}
  L(\bX,\partial_{\bX} S), \partial_{\bX} \phi
  \rangle = \partial_t \phi + (\textbf{a}(t,\bx)-\partial_{\bx}
  S)\partial_{\bx}\phi = \partial_t \phi + \dot
  {\bx}\partial_{\bx}\phi=\frac{D\phi}{Dt} \; ,
\end{equation} 
where $\frac{D}{Dt}$ represents the time derivative along the flow.
This allows us to write the equation $(R_1=0)$ as a simple evolution
equation
\begin{equation}
	\label{eq:r1}
	\frac{D\phi}{Dt} =
        - \left\{ \frac{1}{2}{\rm Tr} \left[ \partial^2_{\bP
              \bP} L(\bX,\partial_{\bX} S)\partial^2_{\bX\bX}S \right]
        +
        {\rm Tr}\left[ \partial^2_{\bX \bP} L \right] \right\} \phi \; .
\end{equation}

To solve this equation we will make use of Liouville's formula, which
states that for a dynamical system 
\begin{equation}
\frac{d\bx}{dt} = f(\bx) \; ,
\end{equation}
and for any $(d\!-\!1)$-parameter family of trajectories
$\bx(t,\boldsymbol{\alpha})$ indexed by 
$\boldsymbol{\alpha} \in \mathbb{R}^{(d-1)}$,  the determinant 
$J(t,\boldsymbol{\alpha}) \equiv
\det\biggl[\frac{\partial\bx(t,\boldsymbol{\alpha})}{\partial
  (t,\boldsymbol{\alpha})}\biggr]$ fulfills 
\begin{equation}
\frac{D\ln J}{Dt}={\rm Tr}\biggl[\frac{d f}{d
  \bx}(\bx(t,\boldsymbol{\alpha}))\biggr] \; .
\end{equation}
(Elements of a demonstration are given in appendix~\ref{app:liouville}
for the sake of completeness.) 
Using the canonical equations we have
\begin{equation}
	\dot{\bX} = \partial_{\bP} L \; .
\end{equation}
Noting that we can write $\bX \equiv (t,\boldsymbol{\xo}(t,\bx_0))$ and having $J$ denote $\det[\partial_{t,\bx_0}\bX]$, Liouville's formula reads

\begin{equation}
	\frac{D \ln(J)}{Dt} = {\rm Tr}\bigl[\partial_\bX(\partial_\bP L)\bigr] = {\rm Tr}[\partial^2_{\bP\bX}L + \partial^2_{\bP\bP}L\partial^2_{\bX\bX}S] \; .
\end{equation}
Hence Eq.~\eqref{eq:r1} becomes
\begin{equation}
	\frac{D \phi}{Dt}  +\frac{1}{2}\frac{D}{Dt}(\ln J)\phi = -\frac{1}{2}{\rm Tr}[\partial^2_{\bX\bP}L]\phi \; ,
\end{equation}
and, multiplying both sides by $\sqrt{J}$,

\begin{equation}
	\frac{D}{Dt}\bigl[\sqrt J\phi\bigr] =
        -\frac{1}{2}{\rm Tr}\left[\partial^2_{\bX\bP}L\right]\sqrt J\phi \; .
\end{equation}
Finally, we have 
\begin{equation}
	\phi(\bx(t,\bx_0)) =
        \frac{\sqrt{J(\bx(t_0,\bx_0))}}{\sqrt{J(\bx(t,\bx_0))}}\phi(\bx(t_0,\bx_0))\exp\biggl(-\frac{1}{2}\int^t_{t_0}
        {\rm Tr}\left[\partial^2_{\bX\bP}L\right]d\tau\biggr) 
\end{equation}
where $\sqrt{J(\bx(t_0,\bx_0)})=1$ and, for $L$ given by
Eq.~\eqref{eq:L-gen},  ${\rm Tr}[\partial^2_{\bX\bP}] =
{\rm div} \, {\bf a}$. In 1d $J$ would simply become
$\partial_{x_0}\xo$, yielding the prefactor in Eq.~\eqref{eq:mv1}.


It is also worth noting that Eq.~\eqref{eq:r1} can be solved in
multiple ways, another possibility would be 
\begin{equation}
	\frac{D}{Dt}\bigl[J\phi\bigr] =
        +\frac{1}{2}{\rm Tr}\left[\left(\partial^2_{\bP^2}L
          \right)\cdot \left(\partial^2_{\bX^2} S \right)
          \right] J\phi \; ,
\end{equation}
implying
\begin{equation}
	\phi(\bx(t,\bx_0)) =
        \frac{J(\bx(t_0,\bx_0))}{J(\bx(t,\bx_0))}\phi(\bx(t_0,\bx_0))\exp\biggl(+\frac{1}{2}\int^t_{t_0}
        {\rm Tr}\left[ \left(\partial^2_{\bP^2}L
          \right)\cdot \left(\partial^2_{\bX^2} S \right) \right]d\tau\biggr) \; .
\end{equation}
Here $J$ serves only as a prefactor; it has no particular physical
meaning, and either expressions cand be used.

\section{Application to the seminar problem}
\label{sec:ill}

For a 1d problem, and writing $\lambda^{-1}\equiv\sigma^2$, the
semiclassical expression for $m$ reads 
\begin{equation}
	m(t,x) = \frac{\Norm}{\sqrt{\partial_{x_0}\xo}}\exp
        \left[\frac{S(t,x)}{\sigma^2} - 
          \frac{1}{2}\int^t_0(\partial_xa)d\tau\right] \; .
	\label{mv2}
\end{equation}
We will use this expression to study the different drift regimes (cf
Eq.~\eqref{eq:a_small_sigma}) presented by the seminar problem for
gaussian initial condition at $t\!=\!0$
\begin{equation}
m_0(x) = \Norm\exp\bigg[-\frac{\mu
	(x_0-\bar x_0)^2}{2\sigma^2}\biggr] =
\Norm\exp\bigg[\frac{S_0(x_0)}{\sigma^2}\biggr] \; ,
\end{equation}
to which through Eq.~\eqref{eq:poxi} we associate the one-parameter
family of intial points in phase space
\begin{equation} \label{eq:family}
\br(x_0) =  ( t\!=\!0, x_0, E_0(x_0), p_0(x_0)) 
\end{equation}
corresponding to 
\begin{equation*}
\begin{aligned}
	p_0(x_0) & = \partial_{x_0} S_0(x_0) = - \mu (x_0 - \bar
        x_0) \; , \\
	E_0(x_0) & = \frac{p^2_0(x_0)}{2} - p_0(x_0) a(x_0,t\!=\!0) \; .
\end{aligned}
\end{equation*}


\subsection{Constant drift}
\label{sec:CstDrift}

Let us start with the simple case of a constant drift $a$ (this would
correspond to regions (0) or (2) in Fig.~\ref{fig:drift}). In order to
obtain the density as expressed in Eq.~\eqref{mv2} there are two terms
we first need to compute, the prefactor $\partial_{x_0}\xo(t,x_0)$
and the action $S(t,x)$. 
To do so we start from the canonical equation of motion
\begin{equation} \label{eq:HamEq-cons-a}
\left\{
\begin{aligned}
&\dot p = - \partial_x L = -p\partial_xa=0\\
&\dot x = \partial_{p} L = a - p  \mbox{   ( = const. along a
  trajectory) . }
\end{aligned} \right .
\end{equation}
For the one-parameter family of trajectories Eq.~\eqref{eq:family},
this leads to
\begin{equation}
\label{eq:conscan}
\left\{
\begin{aligned}
&\po(t,x_0) = \mu(\bar x_0-x_0)\\
&\xo(t,x_0) =x_0+[a-\po(t,x_0)]t  = x_0(1+t\mu) + t(a-\mu \bar x_0) \; .
\end{aligned}
\right. 
\end{equation}
The prefactor is then readily obtained as 
\begin{equation}
\label{pref}
\partial_{x_0}\xo(t,x_0)= 1 +t\mu \; .
\end{equation}
The action is computed noticing that, along the ``center of mass'' trajectory 
$\xo(t,\bar x_0)$, the momentum $\po(t,\bar x_0)$ and energy
$\Eo(t,\bar x_0)$ remain identically zero. Hence, $\mathcal M = \{
(t,\xo(t,x_0),\Eo(t,x_0) \po(t,x_0) \}$ being Lagrangian,
\begin{equation}
	S(t,x) = \int^{x}_{\xo(t,\bar x_0)} p(t,x')dx' 
	\; ,
\end{equation}
$p(t,x)$ being the momentum of the point above $(x,t)$ on $\mathcal M$.
Noting $x_0(t,x)$ the initial position of a trajectory  arriving at 
$x$ at time $t$ (i.e.\ such that $x = \xo(t,x_0))$, the second
equation of \eqref{eq:conscan} gives
\begin{equation}
x_0(t,x)=\frac{x-at+\mu\bar x_0t}{1+\mu t} \; ,
\end{equation}
and the first one
\begin{equation}
\label{constmo}
p(t,x)=
 - \frac{\mu}{1+\mu
  t}\left(x-(\bar x_0+at) \right) \; .
\end{equation}
After integration, this last expression yields, 
\begin{equation}
S(t,x)=-\biggl(\frac{\mu t}{1+\mu t}\biggr)\biggl(\frac{(x-\bar
  x_0-at)^2}{2t}\biggr) \ .
\end{equation}
%
Finally, using Eq.~\eqref{mv2} we have
\begin{equation}
	m(t,x) =
        \sqrt{\frac{\mu}{2\pi\sigma^2}}\frac{1}{\sqrt{1+t\mu}}\exp\biggl[-\biggl(\frac{\mu
          t}{1+\mu t}\biggr)\biggl(\frac{(x-\bar
          x_0-at)^2}{2t\sigma^2}\biggr)\biggr]\; , 
\end{equation}
which turns out to be the {\em exact expression} for the evolution of
a initial Gaussian density in the case of a constant drift.  This is
actually expected since 
going back to the derivation of the semiclassical
approximation, we see that the terms neglected contain only  second (or
higher) order spatial derivative of $a$  which are  identically zero
in the case of a constant drift.

If $\mu \rightarrow \infty$, $m(0,x)
\rightarrow \delta(x-\bar x_0)$, and
\begin{equation}
	m(t,x) \rightarrow G(t,x,\bar x_0) = \sqrt{\frac{1}{2\pi t\sigma^2}}\exp\biggl[-\frac{(x-\bar x_0-at)^2}{2t\sigma^2}\biggr] \; ,
\end{equation}
which indeed is the exact Green function of the Fokker-Planck equation
for a constant drift.

%

\subsubsection*{Absorbing boundary condition at \texorpdfstring{$x=0$}{x=0}} 

To implement the absorbing boundary condition at $x=0$, we follow the
procedure discussed earlier in section~\ref{sec:absorbing} and
construct the ``reflected'' action
\begin{equation} \label{eq:bounced-S}
\tilde{S}(t,x) = S(t,0) + \int_{0}^{x} p_{-}(t,x')dx' \; ,
\end{equation}
where $p_-(t,x)$ is the reflected momentum. 

To compute this quantity, let us note
\[
t_{\rm abs}=\frac{x_0}{\mu(\bar x_0-x_0)-a}
\]
the time at which the trajectory initiated at $x_0$ reaches 0 (and is
thus ``absorbed'').  Since velocity is constant on a given trajectory, we
can express the velocity before the bounce as $\dot x_+(x_0) = -x_0/t_{\rm
  abs}$ and thus just after the bounce as $\dot x_-(x_0) = + x_0/t_{\rm
  abs}$.  Eqs.~\eqref{eq:HamEq-cons-a} then give
\begin{align}
\po_-(t>t_{\rm abs},x_0) & =  a-\frac{x_0}{t_{\rm abs}}  = 2a -
\mu(\bar x_0-x_0) \; , \label{eq:bounced-p}\\
\xo(t>t_{\rm abs},x_0)   & =\frac{x_0}{t_{\rm abs}}(t-t_{\rm abs})
 = - x_0(1+\mu t) - at + \mu \bar x_0 t
\; . \label{eq:bounced-x}
\end{align}
Defining $\tilde x_0(t,x)$ the initial position of a trajectory
arriving at $\xo(t,x_0)=x$ after reflection at $x=0$, we thus have
from Eq.~\eqref{eq:bounced-x}
\begin{equation}
\tilde x_0(t,x)=\frac{\mu t\bar x_0-at-x}{1+\mu t} \; ,
\end{equation}
which inserted into Eq.~\eqref{eq:bounced-p} gives
\begin{equation}
p_-(t,x)=2a-\biggl(\frac{\mu t}{1+\mu t}\biggr)\biggl(\frac{x+\bar
  x_0+at}{t}\biggr) \; . 
\end{equation}
Performing the integral in Eq.~\eqref{eq:bounced-S}, and noting that
the lower bound cancels the term $S(t,x\!=\!0)$,  we thus have
\begin{equation}
\tilde S(t,x) = 2ax-\biggl(\frac{\mu t}{1+\mu
  t}\biggr)\biggl(\frac{(x+\bar x_0+at)^2}{2t}\biggr) \; , 
\end{equation}
giving for the total (incident plus reflected) density
\begin{equation}
\begin{aligned}
m_{\rm tot}(t,x) 
= \sqrt{\frac{\mu}{2\pi\sigma^2}}\frac{1}{\sqrt{1+t\mu}}
&\biggl\{\exp\biggl[-\biggl(\frac{\mu t}{1+\mu
  t}\biggr)\biggl(\frac{(x-\bar x_0-at)^2}{2t\sigma^2}\biggr)\biggr]\\ 
&-\exp\biggl(\frac{2ax}{\sigma^2}\biggr)\exp\biggl[-\biggl(\frac{\mu
  t}{1+\mu t}\biggr)\biggl(\frac{(x+\bar
  x_0+at)^2}{2t\sigma^2}\biggr)\biggr]\biggr\} 
\end{aligned} \; .
\end{equation}
This fulfils the absorbing boundary conditions $m_{\rm tot}(t,0)=0$
and, for the same reason as above, is an exact expression, thus
yielding the exact Green function of the Fokker-Planck equation as
$\mu\rightarrow \infty$.

\subsection{Linear drift}

We will now consider a linear drift $a(x,t)={x}/{(t-T)}$, with $T>t$ the time
at which the seminar begins, associated with region (1) in
Fig.~\ref{fig:drift}. The canonical equations become 
\begin{equation}
\left\{
\begin{aligned}
&\dot x = \partial_{p} L = a - p = \frac{x}{t-T} - p \\
&\dot p = - \partial_x L = -p\partial_xa= -\frac{p}{t-T} \; ,
\end{aligned}
\right. 
\end{equation}
giving 
\begin{equation}
\label{eq:canlin}
\left\{
\begin{aligned}
&\po(t,x_0) = p_0(x_0) \frac{T}{T-t} = \frac{\mu T(\bar x_0-x_0)}{T-t}\\
&\xo(t,x_0) =\frac{x_0(T-t)}{T} - \mu t(\bar x_0-x_0) \; .
\end{aligned}
\right. 
\end{equation}
We thus have $\partial \xo /\partial x_0 = (T-t + \mu tT)/T$, which
together with $\int_0^t (\partial_x a) d\tau = \log[(T-t)/T]$ yieds
for the prefactor to
\begin{equation}
\label{linpref}
\frac{\Norm}{\sqrt{\partial_{x_0}\xo}}\exp
\left[-\frac{1}{2}\int^t_0(\partial_xa)d\tau\right] = 
\sqrt{\frac{\mu}{2\pi\sigma^2}} \sqrt{\frac{T^2}{(\mu t T+T-t)(T-t)}}
\; .
\end{equation}

Turning now to the action, we have from the second equation of
\eqref{eq:canlin}, 
\begin{equation}
x_0(t,x)=(x-\mu t\bar x_0)\frac{T}{T-t+\mu Tt} \; ,
\end{equation}
which, inserted into the first equation of \eqref{eq:canlin}
gives for the  momentum $p(t,x)$,
\begin{equation} \label{eq:p(1)}
p(t,x) = \frac{\mu \, T \, (\bar x_0(T-t)-Tx)}{(T-t) (T-t+\mu Tt)} \; ,
\end{equation}
leading by integration to 
\begin{equation}
\label{eq:linact}
S(t,x)=\int^x_{\xo(t,\bar x_0)} p(t,x')dx'=\frac{\mu
 (\bar x_0(t-T)-Tx)^2}{2(T-t)(T-t+\mu Tt)} \; . 
\end{equation}

Using Eq.~\eqref{mv2}, and computing the reflected action $\tilde
S(t,x)$ following the same procedure as in Section~\ref{sec:CstDrift},
giving 
\begin{equation}
\tilde S(t,x)=\frac{\mu (xT+(T-t) \bar x_0)^2}{2(t-T)(T-t+\mu Tt)} \; ,
\end{equation}
we get for the evolution of a Gaussian initial density with a linear
drift velocity and absorbing boundary conditions at $x\!=\!0$
\begin{equation}
\label{EEL}
\begin{aligned}
m(t,x)& =\sqrt{\frac{\mu}{2\pi\sigma^2}} \sqrt{\frac{T^2}{(\mu t
    T-T-t)(T-t)}} \\
&\biggl\{\exp\biggl[
\frac{\mu (xT-(T-t) \bar  x_0)^2}{2\sigma^2 (t-T)(T-t+\mu Tt)}\biggr]
-\exp\biggl[\frac{\mu (x T + (T-t) \bar x_0)^2}{2\sigma^2
  (t-T)(T-t+\mu Tt)}\biggr]\biggr\}   \; .
\end{aligned} 
\end{equation}
As $\mu \rightarrow \infty$ we recover the Green function of the correponding
Fokker-Planck equation 
\begin{equation}
\label{GFL}
\begin{aligned}
G(t,x,\bar x_0)=\sqrt{\frac{T}{2\pi\sigma^2t(T-t)}}
&\biggl\{\exp\biggl(-\frac{T(x- \frac{T-t}{T}\bar x_0)^2}{2\sigma^2t(T-t)}\biggr)\\
&- \exp\biggl(-\frac{T(x+ \frac{T-t}{T}\bar x_0)^2}{2\sigma^2t(T-t)}\biggr) \biggr\} \; .
\end{aligned}
\end{equation}
Again, because the second $x$ derivative of the drift is zero, expressions \eqref{EEL} and \eqref{GFL} are exact.

\subsection{Coupling the two solutions}
\label{sec:coupling}

We now consider the full problem corresponding to the drift field
Eq.~\eqref{eq:a_small_sigma}, taking into account the possibility that
agents begining in region (0) or (2) (associated with constant drifts
$a^{(0)}$ and $a^{(2)}$) may leak into region (1) (associated with a
linear drift $a(x,t)=x/(t-T)$), and reciprocally. We focus here on
times $t \leq T$ and on the configuration where the agents start their
diffusion in region (1), which is the one of interest from the point
of view of mean field games.  Corresponding expressions for a group
of agents initially located in region (2) are given in
appendix~\ref{app:Coupling}.

We begin by defining $x^{*(n)}(x_0)$, $p^{*(n)}(x_0)$ and
$t^{*(n)}(x_0)$, ($n = 0,2$), the position, impulsion and time at
which a trajectory intiated at ${\bf r}(x_0)$ (cf
Eq.~\eqref{eq:family}) crosses the boundary between regions (1) and
$(n)$. Using Eq.~\eqref{eq:canlin} together with the fact that the
boundary is the $x=a^{(n)}(t-T)$ straight line, we may write
\begin{equation}
x^{*(n)}(x_0) = a^{(n)}(t^{*(n)}-T) = x_0\frac{(T-t^{*(n)})}{T} - \mu(\bar x_0 -x_0)t^{*(n)}(x_0) \; .
\end{equation}
We then compute $t^{*(n)}$ by inverting this last equation and obtain
$p^{*(n)}$ inserting this newly found $t^{*(n)}$ expression in
Eq.~\eqref{eq:canlin} 
\begin{equation}
\label{star}
\left\{
\begin{aligned}
& t^{*(n)}(x_0) = T \biggl[1-\frac{\mu T(\bar x_0 -x_0)}{a^{(n)}T + x_0 + \mu T(\bar x_0 -x_0)}\biggr]\\
& p^{*(n)}(x_0) = \frac{a^{(n)}T+\mu T(\bar x_0-x_0)+x_0}{T}
\end{aligned}
\right.  \qquad .
\end{equation}

Before the crossing ($t<t^{*(n)}$) the agents do not feel the effects of
the drift change, and their trajectories remain the same as in
Eq.~\eqref{eq:canlin}. In region (1), 
($x^{*(0)}<x<x^{*(2)}$), the prefactor is thus obtained as 
Eq.~\eqref{linpref} and the action as Eq.~\eqref{eq:linact}. We will now
focus on the expression of the density after the crossing, the
complete solution being simply obtained by patching the linear and the
leaking densities. 

Using the canonical equations in the region in which the agents are
leaking, we have for $t>t^{*(n)}$,
\begin{equation}
\label{cancoup}
\left\{
\begin{aligned}
& \xo^{(n)}(t,x_0) =x^{*(n)} + (a^{(n)}-p^{*(n)})(t-t^{*(n)})=\x_0\frac{T-t}{T}-t\mu(\bar x_0-x_0)\\
& \po^{(n)}(t,x_0)=p^{*(n)}
\end{aligned}
\right. \; \; .
\end{equation}
Let $x_0(t,x)$ be the initial position of a trajectory arriving at $x$
at time $t$ (thus $\xo_n(t,x_0)=x$), $t^{*(n)}(t,x)$ the time at which
this trajectory crosses the boundary between the two regions, and
$p^{*(n)}(t,x)$ the momentum at the crossing
\begin{equation}
\label{starfun}
\left\{
\begin{aligned}
&x_0(t,x)=\frac{T(x-\mu t\bar x_0)}{T-t-\mu Tt} \\
& t^{*(n)}(t,x) = T \biggl\{1-\frac{\mu T\bigl[\bar x_0 -x_0(t,x)\bigr]}{aT + x_0(t,x) + \mu T(\bar x_0 -x_0(t,x))}\biggr\}\\
& p^{*(n)}(t,x) = \frac{aT+\mu T\bigl[\bar x_0-x_0(t,x)\bigr]+x_0(t,x)}{T}\\
\end{aligned}
\right. \; \; .
\end{equation}
We may now compute the prefactor
\begin{equation}
\begin{aligned}
\frac{\Norm}{\sqrt{\partial_{x_0}\xo_n}}&\exp
\left[-\frac{1}{2}\int^{t^*(t,x)}_0(\partial_x a^{(1)}) d\tau\right]=\\
&\sqrt{\frac{\mu}{2\pi\sigma^2}\frac{T}{T-t+\mu Tt} \frac{\mu T(\bar
    x_0-x)+x+a^{(n)}(T-t+\mu Tt)}{\mu T(\bar x_0-x)-\mu t\bar x_0}} \; ,
\end{aligned} 
\end{equation}
and the action
\begin{equation}
\begin{aligned}
&S^{(n)}_{\rm{leak}}(t,x) = \int^{a^{(n)}(t-T)}_{\bar x} p^{(1)}(t,x') dx' + \int^x_{a^{(n)}(t-T)}
p^{*(n)}(t,x') dx' \\
&= \frac{-(a^{(n)})^2(t-T)(T-t+\mu Tt)+2a^{(n)}(T-t+\mu Tt)x+(1-\mu
  T)x^2+2\mu Tx\bar x_0+\mu(t-T)\bar x_0^2}{2(T-t+\mu tT)} \; ,
\end{aligned} 
\end{equation}
with $p^{(1)}$ given by Eq.~\eqref{eq:p(1)}.  We note that if
both $x$ and $\bar x_0$ belong to the boundary between region (1) and
region (n), the prefactor diverges because of diffraction effects that
should be treated specifically.

The reflected action is computed through the usual
procedure, but, this time, taking into account that the reflected
trajectory may also transit from a region to an other 
\begin{equation}
\label{sleak}
\begin{aligned}
\tilde
S_{\rm{leak}}(t,x)=&S^{(n)}_{\rm{leak}}(t,0)+\int_{0}^{\min[x;a^{(1)}(t-T)]}
p^{*(0)}_{-}(t,x')dx'\\
&+\int_{a^{(1)}(t-T)}^{\min\left[\max[x;a^{(1)}(t-T)];a^{(2)}(t-T)\right]}p^{(1)}_{-}(t,x')dx'\\
&+\int_{a^{(2)}(t-T)}^{\max[x,a^{(2)}(t-T)]}p^{*(2)}_{-}(t,x')dx'
\; ,
\end{aligned} 
\end{equation}
with $p^{*(n)}_{-}$ the reflected leaking momentum in region (n)
  and $p^{(1)}_{-}$ the reflected linear drift momentum. Complete,
explicit, expressions are given in appendix~\ref{app:Coupling} (cf 
Eqs.~\eqref{near}, \eqref{mid} and \eqref{far}).  However the
contribution of reflected trajectories decay exponentially away from
the absorbing boundary $x=0$. Assuming $t \leq T$ as we do here, this
implies that unless $t\approx T$, we can assume the contribution of
reflected trajectories are important only when they are still in
region (0), and the reflected action can be approximated as
\begin{equation}
\begin{aligned}
\tilde S_{\rm{leak}}(t,x) = 2a^{(0)} x -\frac{1}{2(T-t+\mu tT)}&\biggl[-(a^{(0)})^2(t-T)(T-t+\mu Tt)+2a^{(0)}(T-t+\mu Tt)x\\
&+(1-\mu T)x^2+2\mu Tx\bar x_0+\mu(t-T)\bar x_0^2\biggr] \; .
\end{aligned}
\end{equation}
We can show that, for this specific drift field, the reflected prefactor is the same as the direct one. Eventually, using Eq.\eqref{mv2}, we  have
\begin{equation}
\begin{aligned}
m_{\rm{leak}}(t,x)=&\sqrt{\frac{\mu}{2\pi\sigma^2}\frac{T}{T-t+\mu
    Tt}\frac{\mu T(\bar x_0-x)+x+a^{(n)}(T-t+\mu Tt)}{\mu T(\bar
    x_0-x)-\mu t\bar x_0}}\biggl\{\\ 
&\exp\biggl(\frac{S_{\rm{leak}}(t,x)}{\sigma^2}\biggr)-\exp\biggl(\frac{\tilde
  S_{\rm{leak}}(t,x)}{\sigma^2}\biggr)\biggr\}  \; .
\end{aligned}
\end{equation}

\begin{figure} [h!]
	\includegraphics[scale=0.41]{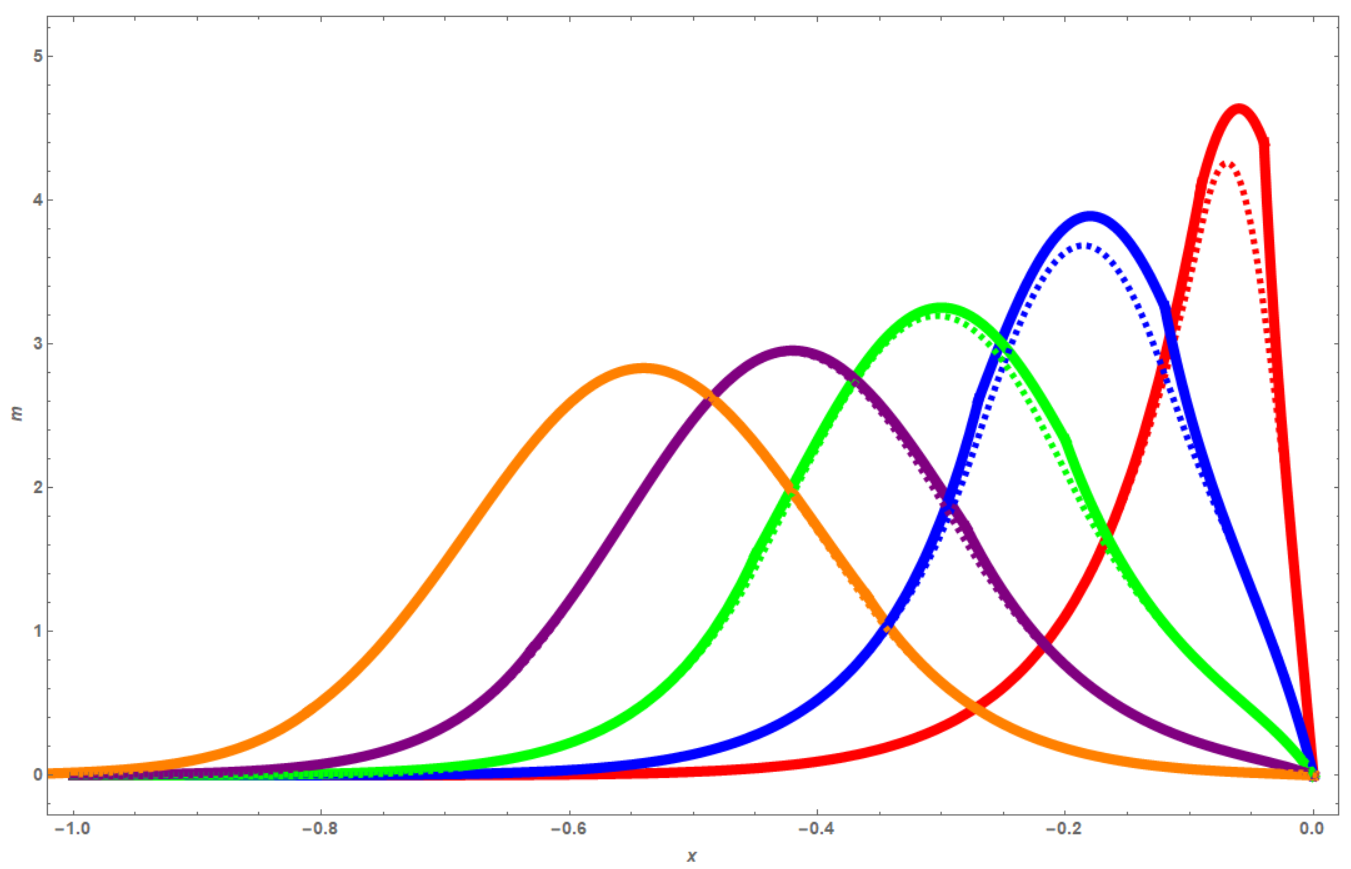}
	\caption{Spatial distribution of the agents, dashed lines show the numerical solution while solid lines show the approximation. From left to right, $K=0.19$, $K=0.24$, $K=0.33$, $K=0.56$, $K=1.67$. In this case $T=2$, $a^{(0)}=0.4$, $a^{(2)}=0.9$, $\sigma=0.2$, $\bar x_0 =1.2$ and $\mu=10^6$.}
	\label{fig:numanal1}
\end{figure}

\begin{figure} [h!]
	\includegraphics[scale=0.39]{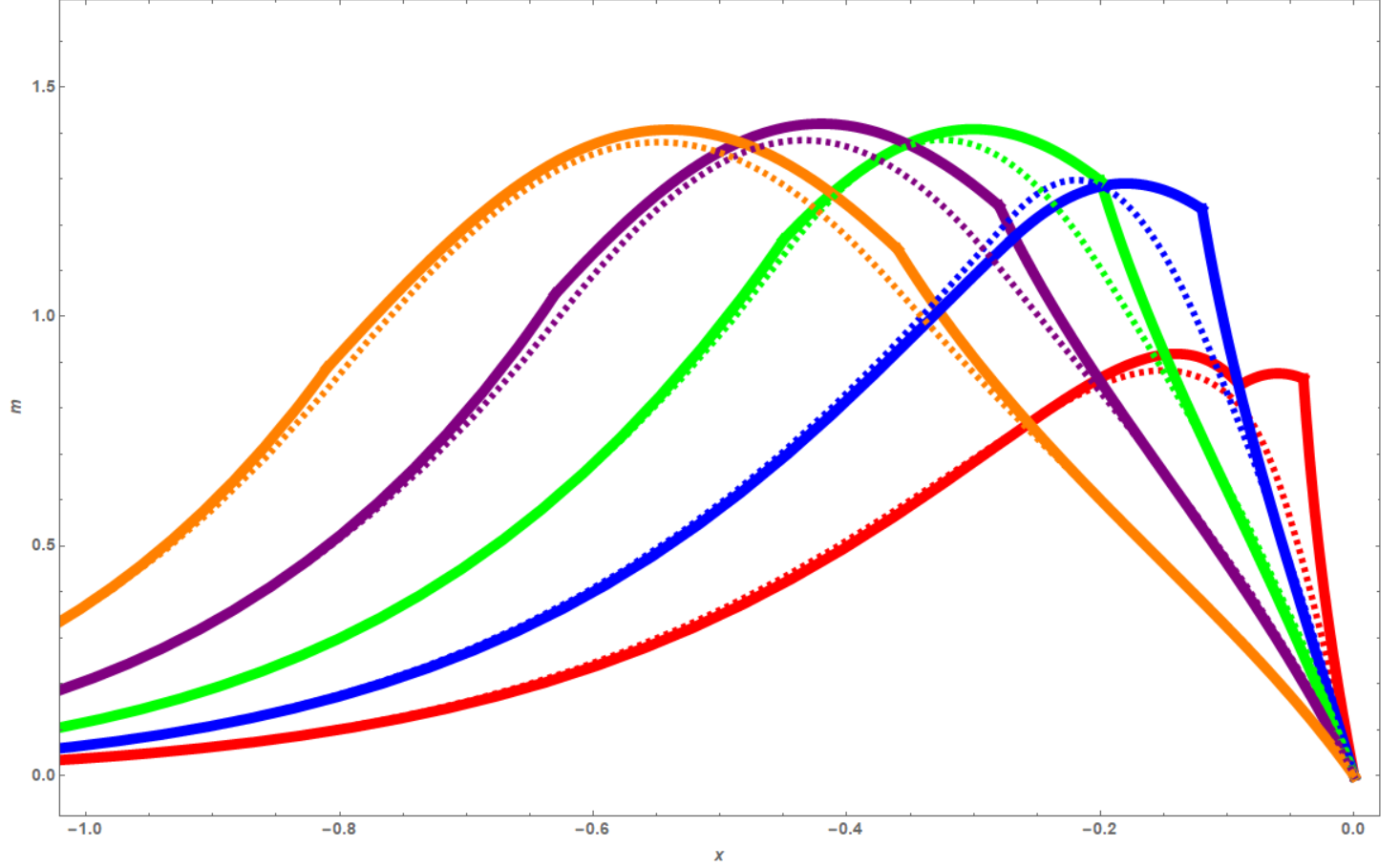}
	\caption{Spatial distribution of the agents, dashed lines show the numerical solution while solid lines show the approximation. From left to right, $K=0.74$, $K=0.95$, $K=1.33$, $K=2.22$, $K=6.66$. In this case $T=2$, $a^{(0)}=0.4$, $a^{(2)}=0.9$, $\sigma=0.4$, $\bar x_0 =1.2$ and $\mu=10^6$.}
	\label{fig:numanal2}
\end{figure}

Contrarily to constant and linear drifts which represent non-generic
cases for which the WKB expression is exact, the above result is an
approximation valid only in the semiclassical regime of small
$\sigma$'s.  To be a bit more quantitative, we thus introduce the
dimensionless parameter $K$ defined as the ratio between the drift
time $\tau_{\rm{drift}}={\xo(t,\bar x_0)}/{a}$, the time needed to get
from $x=\xo(t,\bar x_0)$ to the location of the absorbing boundary
condition $x=0$ at speed $a$, and the diffusion time $\tau_{\rm
  {diffusion}}={\xo^2(t,\bar x_0)}/{\sigma^2}$, time it would take to
a purely diffusive process to spread the density from its center in
$x=\xo(t,\bar x_0)$ to $x=0$.  Thus
\begin{equation}
K=\frac{\tau_{\rm{drift}}}{\tau_{\rm {diffusion}}} = \biggl|\frac{\sigma^2}{a\xo(t,\bar
x_0)}\biggr| \propto \sigma^2
\; .  
\end{equation}
The ``small noise'' [semiclassical] regime can be therefore characterized
by $K \ll 1$, and the large noise regime by $K \gg 1$.  Note that $K$
usually depends on time. Fig.~\ref{fig:numanal1} shows a comparison
between a numerical solution and the semiclassical approximation for
different small values of $K$, fixing $\sigma$ and varying $t$.
As we can see the semiclassical approximation is almost indistinguishable
from the numerical solution up to $K=0.33$ and remains good for $K$
slightly greater than one even if we can observe small
  discrepancies. Looking at larger values of $\sigma$ (and 
hence $K$), cf. Fig.~\ref{fig:numanal2}, we see that even for the
largest value of $K$ considered ($K=6.66$), the agreement is still rather good although the difference with
the exact result becomes more significant. 
The fact that the source of errors in the
semiclassical treatment is generated only at the boundaries between the
various regions explains the effectiveness of the approximation in
this particular setup.

\section{Conclusion}
\label{sec:conclusion}

 In this paper we proposed a new take on the WKB approximation scheme
 to study the Fokker-Planck equation. This approach, based on Maslov's
 geometric perspective, offers what we think to be a transparent way of tackling the Fokker-Planck
 equation, which  we illustrated here on a problem motivated by a simple toy model of
 mean field games theory.

 As stressed in the introduction, we have addressed here only a very
 small part of the program which would consist in providing a ``ray
 theory'' of mean field games in the small but non zero-noise limit.
 This program would involve a few steps (to start with a ray theory of
 the Hamilton-Jacobi-Bellman equation and then dealing with the
 coupling between the two) which are significantly more involved.  We
 leave these for future research, but we are convinced  that
 the WKB approach we propose provide a sound start for this program.

\appendix

\section{Method of characteristics}
\label{app:characteristics}

The method of characteristics is typically used to solve first-order
partial differential equations. It aims to reduce a PDE to a family of
ODEs that can be easily integrated.  A rather complete discussion of
this method can be found for instance in  chapter II of
\cite{CourantMathPhy}.

%
%
%
%
%
%
%
%
%
In the particular case of the  Hamilton-Jacobi equation
\begin{equation}
 \partial_tS + a\partial_x S - \frac{1}{2}(\partial_x S)^2 = 0 \; ,
\end{equation}
it is however extremely straigtforward to check that the action
defined by Eq.~\eqref{eq:action} is a solution.  Indeed, using the
least action principle, one has that for any $X=(x,t)$, $\partial_x
S=p$ and $\partial_t S =E$, with $p$ and $E$ the momentum and energy
of the trajectory reaching $x$ at time $t$.  Since all the
trajectories involved have to fulfill the compatibility condition
Eq.~\eqref{eq:compatibility}, this one reads $L(x,t;\partial_x
S,\partial_t S) = 0$, which is precisely the Hamilton-Jacobi equation.

\section{Liouville's formula}
\label{app:liouville}

For completeness, in this appendix, we provide a brief derivation of
the Liouville formula used in Section~\ref{sec:proof}, as presented in
\cite{SmirnovHMIV}. We consider a dynamic system described by
\begin{equation}
	\frac{d\bx}{dt} = f(\bx) \quad (x \in \mathbb{R}^d) \; ,
\end{equation}
and consider a $(d-1)$-family of trakectories
$\bx(t,\boldsymbol{\alpha})$ indexed by $\boldsymbol{\alpha} \in
\mathbb{R}^{(d-1)}$ . Defining
$J(t,\boldsymbol{\alpha}) \equiv
\det\biggl[\frac{\partial\bx(t,\boldsymbol{\alpha})}{\partial
  (t,\boldsymbol{\alpha})}\biggr]$, the Liouville's formula states
that :
\begin{equation}
	\frac{d\ln J}{dt}={\rm Tr}\biggl[\frac{\partial f}{\partial
          \bx}(\bx(t,\boldsymbol{\alpha}))\biggr] \; .
\end{equation}
	
\subsection*{Derivation}
	
Let  $A$ a $d \times d$ matrix.  We have $\det A = \exp[{\rm Tr} \ln
A]$, and thus 
\begin{equation}
	\frac{d(\ln \det A)}{dt} =   \frac{d ({\rm Tr} \ln A)}{dt} \; .
\end{equation}
Now, for any function $g$ of $A$, writting  $g(A) = \sum_{n} g_nA^n$
and using the cyclicity of the trace we have
\begin{equation}
	\frac{d ({\rm Tr} g(A))}{dt} 
	={\rm Tr}\biggl[g'(A)\frac{dA}{dt}\biggr] \; .
\end{equation}
Thus, if $A \equiv \frac{\partial \bx}{\partial(t,
  \boldsymbol{\alpha})}$ and $J(t,\boldsymbol{\alpha}) \equiv \det A$,
we have 
\begin{equation}	
		\frac{d\ln J}{dt}= {\rm Tr} {A^{-1} \frac{d A}{dt}}
                \; .
\end{equation}
Noting that here the total derivative $\frac{d}{dt}$ is the
same as the partial derivative $\partial_t$ taken at contant
$\boldsymbol{\alpha}$, one furthermore has
\begin{equation}	
	 \frac{d A}{dt} = \frac{\partial^2 \bx (t,\boldsymbol{\alpha} ) }{\partial t \partial
           (t,\boldsymbol{\alpha} )}
         =  \frac{\partial f(\bx (t,\boldsymbol{\alpha} ))}{ \partial
           (t,\boldsymbol{\alpha})}
                \; .
\end{equation}
Thus
\begin{equation}	
  \frac{d\ln J}{dt} = {\rm Tr}
  \left[\frac{\partial (t,\boldsymbol{\alpha})}{\partial \bx}  
   \frac{\partial f(\bx (t,\boldsymbol{\alpha} ))}{ \partial
           (t,\boldsymbol{\alpha})}  \right]= 
{\rm Tr}\biggl[\frac{\partial
                  f}{\partial \bx}(\bx(t,\boldsymbol{\alpha}))\biggr]
                \; .
\end{equation}

\section{Coupling the two solutions}
\label{app:Coupling}
This appendix aims at addressing what we left out of
\ref{sec:coupling} for the sake of succinctness. We will first provide
explicit expressions for the reflected action Eq.~\eqref{sleak}, then
we will dicuss the configuration where the agents begin in a constant
drift region.

\subsection*{Explicit expression of the reflected action}

Recalling Eq.~\eqref{sleak}
\begin{equation}
	\begin{aligned}
		\tilde
		S_{\rm{leak}}(t,x)=&S^{(n)}_{\rm{leak}}(t,0)+\int_{0}^{\min[x;a^{(1)}(t-T)]}
		p^{*(0)}_{-}(t,x')dx'\\
		&+\int_{a^{(1)}(t-T)}^{\min\left[\max[x;a^{(1)}(t-T)];a^{(2)}(t-T)\right]}p^{(1)}_{-}(t,x')dx'\\
		&+\int_{a^{(2)}(t-T)}^{\max[x,a^{(2)}(t-T)]}p^{*(2)}_{-}(t,x')dx'
		\; ,
	\end{aligned} 
\end{equation}
there are three domains in which $\tilde S_{\rm{leak}}(t,x)$ takes slightly diffrent expressions.
\begin{itemize}
	\item $x<a^{(1)}(t-T)$
	
	\begin{equation}
		\label{near}
		\begin{aligned}
			\tilde S_{\rm{leak}}(t,x) = 2a^{(0)}x -\frac{1}{2(T-t+\mu tT)}&\left[-(a^{(0)})^2(t-T)(T-t+\mu Tt)+2a^{(0)}(T-t+\mu Tt)x\right.\\
			&\left.+(1-\mu T)x^2+2\mu Tx\bar x_0+\mu(t-T)\bar x_0^2\right]
		\end{aligned} \; .
	\end{equation}
	\item $a^{(1)}(t-T)<x<a^{(2)}(t-T)$
	\begin{equation}
		\label{mid}
		\begin{aligned}
			\tilde S_{\rm{leak}}(t,x) = \frac{1}{2(T-t+\mu tT)(t-T)}&\left[\mu(T^2x^2-2T(T-t)(2a^{(0)}(T-t)+x)x_0\right.\\
			&\left.+(t-T)^2x_0^2\right]
		\end{aligned} \; .
	\end{equation}
	\item $x>a^{(2)}(t-T)$
	\begin{equation}
		\label{far}
		\begin{aligned}
			\tilde S_{\rm{leak}}(t,x) = \frac{1}{2(T-t+\mu tT)}&\left[(a^{(2)})^2(t-T)(T-t+\mu t T)-2a^{(2)}(T-t+\mu tT )x\right.\\
			&\left.+(\mu T-1)x^24a^{(2)}\mu T(T-t)x_0\right.\\
			&\left. +2\mu Txx_0+\mu(T-t)(4a^{(1)}T-x_0)x_0\right]
		\end{aligned} \; .
	\end{equation}
\end{itemize}
\begin{figure} [h!]
	\includegraphics[scale=0.40]{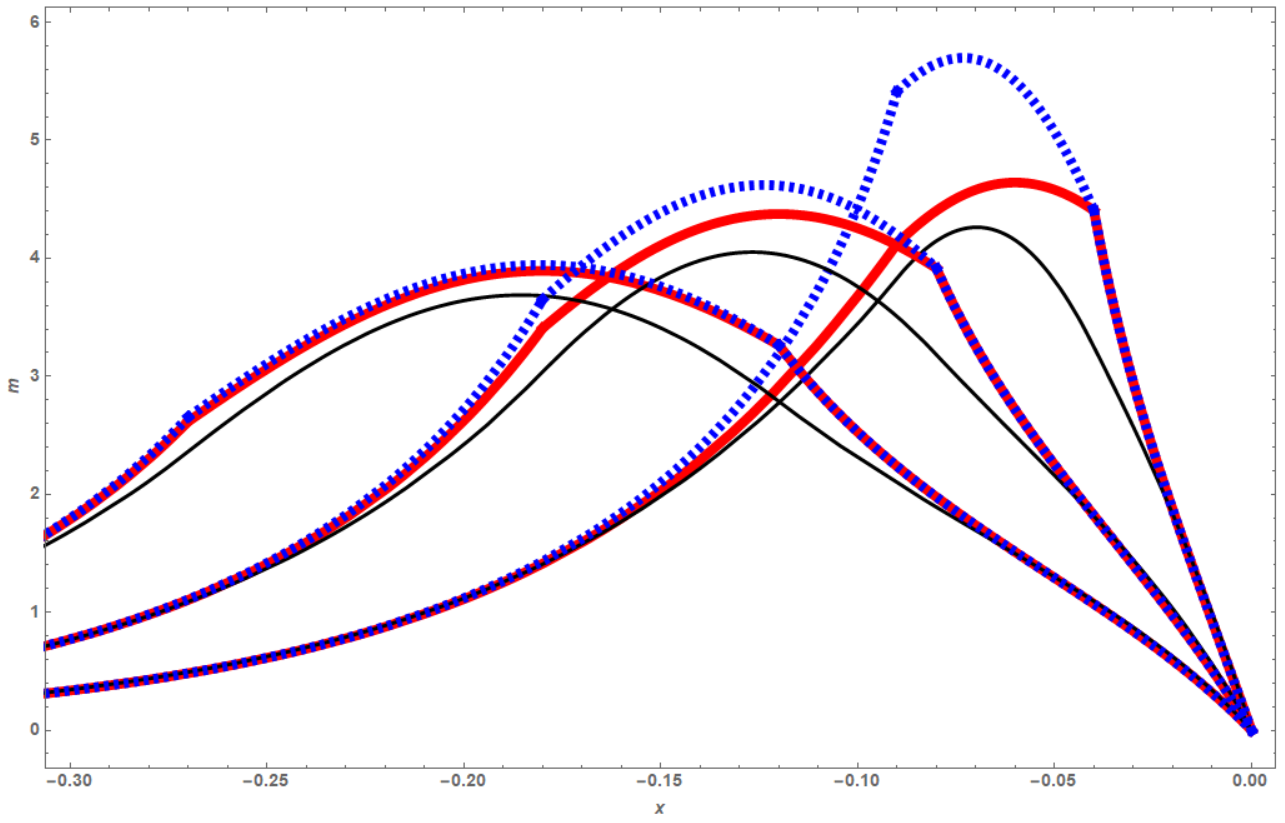}
	\caption{Spatial distribution of the agents, the slim line
          represents the numerical solution, the thick straight line
          the approximation using Eq.~\eqref{sleak} and the thick dashed
          line the approximation using only Eq.~\eqref{near}. In this
          case $T=2$, $a^{(0)}=0.4$, $a^{(2)}=0.9$, $\sigma=0.2$,
          $\bar x_0 =1.2$ and $\mu=10^6$. From left to right, $t=1.9$,
          $t=1.8$, $t=1.7$. }
	\label{fig:appc}
\end{figure}
However, as mentioned in section \ref{sec:coupling}, Eq.~\eqref{sleak} can be approximated using only Eq.~\eqref{near}. This is shown in Fig.~\ref{fig:appc} where the results of the two approximations, although obviously different for $t\approx T$ become more and more similar the smaller $t$ gets.

\subsection*{Leak from a constant to a linear drift region}
We begin, as in Section~\ref{sec:coupling}, by computing the position,
time and momentum of the agents as they cross the boundary between an
region of constant drift $a^{(n)}$ and region (1). Keeping the same
notations and using the same method as earlier we have 
\begin{equation}
\left\{
\begin{aligned}
& x^{*(n)}(x_0)=a^{(n)}(t^{*(n)}-T)=x_0(1+t^{*(n)}\mu)+t^{*(n)}(a^{(n)}-\mu x_0)\\
& t^{*(n)}(x_0) = \frac{a^{(n)}T+x_0}{\mu (\bar x_0 -x_0)}\\
& p^{*(n)}(x_0) =\mu (\bar x_0-x_0)\\
\end{aligned}
\right. \; .
\end{equation}
Using the canonical equations in region (1), we compute for $t>t^{*(n)}(x_0)$ 
\begin{equation}
\label{cancoup-app}
\left\{
\begin{aligned}
& \po^{(n)}(t,x_0)=\frac{aT+x_0-\mu T(\bar x_0-x_0)}{t-T}\\
& \xo^{(n)}(t,x_0)=at+x_0-\mu t(\bar x_0-x_0)\\
& x_0(t,x)=\frac{x-at+\mu t \bar x_0}{1+\mu t}
\end{aligned}
\right. \; ,
\end{equation}
from which we get the prefactor
\begin{equation}
\begin{aligned}
\frac{\Norm}{\sqrt{\partial_{x_0}\xo_n(t,x_0)}}&\exp
\left[-\frac{1}{2}\int^{t}_{t^{*(n)}(t,x)}(\partial_xa^{(1)})d\tau\right]=\\
&\sqrt{\frac{\mu}{2\pi\sigma^2(1+t\mu)}\frac{\mu (t-T)(at-x+\bar x_0)}{a(T-t)+x+\mu Tx-\mu(T-t)\bar x_0}}
\end{aligned} \; ,
\end{equation}
and the action
\begin{equation}
\begin{aligned}
S^{(n)}_{\rm{leak}}(t,x) &= \int^{a^{(n)}(t-T)}_{\xo(t,\bar x_0)} p^{(n)}(t,x') dx' + \int^x_{a^{(n)}(t-T)}
p^{*(1)}(t,x') dx'\\
\end{aligned} \; ,
\end{equation}
with $p^{(n)}$ the constant drift momentum of region (n) given by
Eq.~\eqref{constmo} and $p^{*(1)}$ the leaking momentum in region (1)
obtained by inserting the third equation of Eqs.~\eqref{cancoup-app} into the second, yielding
\begin{equation}
\begin{aligned}
S^{(n)}_{\rm{leak}}(t,x)
&= \frac{1}{2(1+\mu t)(t-T)}\left[ -a^{(n)2}(t-T)(T-t+\mu tT)+x^2(1+\mu T)\right.\\
&\left.+2\mu(t-T)x\bar x_0+\mu (T-t)x_0^2-2a^{(n)}(t-T)(x+\mu t\bar x_0)\right]
\end{aligned} \; .
\end{equation}
In the case where agents begin in region (2), they may diffuse up to regon (0), using, once again the same scheme, we compute the new prefactor 
\begin{equation}
\begin{aligned}
\frac{\Norm}{\sqrt{\partial_{x_0}\xo_0(t,x_0)}}&\exp
\left[-\frac{1}{2}\int^{t^{*(2)}(t,x)}_{t^{*(1)}(t,x)}(\partial_xa^{(1)})d\tau\right]=\\
&\sqrt{\frac{\mu}{2\pi\sigma^2(1+t\mu)}\frac{\mu (a^{(2)}t-x+\bar x_0)}{a^{(0)}-a^{(2)}+a^{(0)}\mu t+\mu(\bar x_0-x)}}
\end{aligned} \; ,
\end{equation}
and the new action
\begin{equation}
\begin{aligned}
S^{(0)}_{\rm{leak}}(t,x)
&= \int^{a^{(2)}(t-T)}_{\xo(t,\bar x_0)} p^{(n)}(t,x') dx' + \int^{a^{(0)}(t-T)}_{a^{(2)}(t-T)}
p^{*(1)}(t,x') dx'+\int^x_{a^{(0)}(t-T)}
p^{*(0)}(t,x') dx'\\
&=-\frac{1}{2(1+\mu t)}\left[a^{(2)2}(3+\mu t)(t-T)-a^{(0)2}(T-t+\mu tT)\right.\\
&\left.-2a^{(2)}x+2a^{(0)}\left(-a^{(2)}(2+\mu t)(t-T)+x+\mu t(x-\bar x_0)\right)-\mu(x-\bar x_0)^2\right]
\end{aligned} \; .
\end{equation}
Finally the reflected action is computed as
\begin{equation}
\begin{aligned}
\tilde
S_{\rm{leak}}(t,x)=&S^{(0)}_{\rm{leak}}(t,0)+\int_{0}^{\min[x;a^{(1)}(t-T)]}
p^{*(0)}_{-}(t,x')dx'\\
&+\int_{a^{(1)}(t-T)}^{\min\left[\max[x;a^{(1)}(t-T)];a^{(2)}(t-T)\right]}p^{*(1)}_{-}(t,x')dx'\\
&+\int_{a^{(2)}(t-T)}^{\max[x,a^{(2)}(t-T)]}p^{(2)}_{-}(t,x')dx'
\; ,
\end{aligned} 
\end{equation}
that we approximate, as in Section~\ref{sec:coupling}, as
\begin{equation}
\begin{aligned}
\tilde
S_{\rm{leak}}(t,x)
&=2a^{(0)}x+\frac{1}{2(1+\mu t)}\left[a^{(2)2}(3+\mu t)(t-T)-a^{(0)2}(T-t+\mu tT)\right.\\
&\left.-2a^{(2)}x+2a^{(0)}\left(-a^{(2)}(2+\mu t)(t-T)+x+\mu t(x-\bar x_0)\right)-\mu(x-\bar x_0)^2\right]
\end{aligned} \; .
\end{equation}

%

\end{document}